\documentclass[sn-nature]{sn-jnl}

\usepackage{graphicx}%
\usepackage{multirow}%
\usepackage{amsmath,amssymb,amsfonts}%
\usepackage{amsthm}%
\usepackage{mathrsfs}%
\usepackage[title]{appendix}%
\usepackage{xcolor}%
\usepackage{textcomp}%
\usepackage{manyfoot}%
\usepackage{booktabs}%
\usepackage{algorithm}%
\usepackage{algorithmicx}%
\usepackage{algpseudocode}%
\usepackage{listings}%
\usepackage{romannum}%
\usepackage{siunitx}%
\usepackage{braket}%
\usepackage{geometry}%

\newgeometry{left=3.5cm, right=3.5cm}

\begin{document}

\title[Article Title] {Imaging magnetic flux trapping in lanthanum hydride using diamond quantum sensors}
%\title[Article Title]{Imaging magnetism in LaH$_x$ using nitrogen-vacancy centers in diamond}
%\title[Article Title] {Imaging magnetic field screening and flux trapping in lanthanum hydride using diamond quantum sensors}

\author[1]{\fnm{Yang} \sur{Chen}}
\equalcont{These authors contributed equally to this work.}

\author[1,2]{\fnm{Junyan} \sur{Wen}}
\equalcont{These authors contributed equally to this work.}

\author[1,2]{\fnm{Ze-Xu} \sur{He}}
\equalcont{These authors contributed equally to this work.}

\author[3]{\fnm{Jing-Wei} \sur{Fan}} 

\author[1,2]{\fnm{Xin-Yu} \sur{Pan}}

\author[4]{\fnm{Cheng} \sur{Ji}}
\author[4]{\fnm{Huiyang} \sur{Gou}}

\author*[1,2]{\fnm{Xiaohui} \sur{Yu}} \email{yuxh@iphy.ac.cn}

\author*[1]{\fnm{Liucheng} \sur{Chen}} \email{chenliucheng@iphy.ac.cn}

\author*[1,2]{\fnm{Gang-Qin} \sur{Liu}} \email{gqliu@iphy.ac.cn}

\affil[1]{\orgdiv{Beijing National Laboratory for Condensed Matter Physics and Institute of Physics}, \orgname{Chinese Academy of Sciences}, \orgaddress{\city{Beijing}, \postcode{100190}, \country{China}}}

\affil[2]{\orgdiv{School of Physical Sciences}, \orgname{ University of Chinese Academy of Sciences}, \orgaddress{\city{Beijing}, \postcode{100190}, \country{China}}}

%\affil[3]{\orgdiv{CAS Center of Excellence in Topological Quantum Computation}, \orgaddress{\city{Beijing}, \postcode{100190}, \country{China}}}

\affil[3]{\orgdiv{Department of Physics, Hefei University of Technology}, \orgaddress{\city{Hefei}, \postcode{230009}, \country{China}}}

\affil[4]{\orgdiv{Center for High Pressure Science and Technology Advanced Research}, \orgaddress{\city{Beijing}, \postcode{100190}, \country{China}}}

\abstract{Lanthanum hydride has attracted significant attention in recent years due to its signatures of superconductivity at around 250 K \cite{LaH2019Nature, LaH2019PRL}. However, the megabar pressures required for synthesize and maintain its state present extraordinary challenges for experiments, particularly in characterizing its Meissner effect \cite{Mag2023NP, H_SC_2025NP}. The nitrogen-vacancy (NV) center in diamond has emerged as a promising quantum probe to address this problem \cite{NVHP_2014PRL, hsieh2019imaging, lesik2019magnetic, yip2019measuring}, but a gap remains between its working pressure and the pressure required to study the superconducting state of lanthanum hydride \cite{NVHP2022CPL, NVHP2023PRB, NVHP2024NC, bhattacharyya2024imaging}. In this work, using neon gas as the pressure transmitting medium, the working pressure of NV centers is extended to nearly 200 GPa. This quantum probe is then applied to study the Meissner effect of a LaH$_{9.6}$ sample, synthesized by laser heating ammonia borane and lanthanum. A strong magnetic shielding effect is observed, with the transition temperature beginning at around 180 K and completing at 220 K.
%After zero-field cooling to 100 K and measuring under an external magnetic field of 105 G, a strong magnetic shielding effect is observed.
%(up to\textcolor{red}{70\%} at the center) and flux concentration at the edge observed simultaneously in the sample. 
In addition, magnetic field imaging after field cooling reveals strong flux trapping and significant inhomogeneities within the sample. Our work provides compelling evidence for superconductivity in lanthanum hydride and highlights the importance of spatially resolved techniques in characterizing samples under ultrahigh pressure conditions.
}

\maketitle

\subsection*{Introduction}
In recent years, compressed hydrides have attracted significant attention as promising candidates for room-temperature superconductors due to the high-frequency phonons of hydrogen and the chemical precompression from the hydrogen sublattice \cite{ashcroft1968metallic, CeH2012PNAS, HS32015Nature}. A wide range of superconducting hydrides have been successfully synthesized in experiments, spanning binary, ternary, and even quaternary systems, such as H$_3$S \cite{HS32015Nature}, LaH$_{10}$ \cite{LaH2019Nature, LaH2019PRL}, CaH$_6$ \cite{CaH2022PRL}, YH$_9$ \cite{YH2021NC}, LaBeH$_8$ \cite{LaBeH8_2023PRL}, (Y,Ce)H$_9$ \cite{CeYH9_2024NC}, and (La,Y,Ce)H$_{10}$ \cite{LaYCeH_2024JACS}. However, there is growing skepticism in this field about whether high-pressure hydrides host superconductivity \cite{Hirsch2022, Hirsch2024NSR}. The main challenge arises from the fact that only limited  characterization methods are available under megabar pressures, especially for  measuring magnetic properties. The widely used SQUID and a.c. magnetic susceptibility methods measure the magnetic response of the entire diamond anvil cell (DAC) \cite{Mag2019NSR, Mag2022NC, Mag2023NP}. Consequently, the weak diamagnetic signal from the tiny sample is easily overwhelmed by the substantial background from the DAC. Furthermore, these methods generally provide global information about the sample, lacking spatial resolution, which is crucial for understanding the homogeneity of the hydride sample.

With exceptional magnetic field sensitivity, submicron spatial resolution, and notably wide working ranges, quantum sensing using diamond nitrogen-vacancy (NV) centers offers an ideal solution to this problem. In-situ magnetic field imaging can be achieved by combining optically detected magnetic resonance (ODMR) with DAC-based high-pressure techniques, and quantum control of NV spins has been demonstrated \cite{NVHP_2014PRL, NVHP2019CPL, hsieh2019imaging, lesik2019magnetic, yip2019measuring}. Subsequently, the working pressure of NV centers has been increased to around 140 GPa \cite{NVHP2022CPL, NVHP2023PRB, NVHP2024NC, bhattacharyya2024imaging}, and recently this method has been used to probe the Meissner effect in CeH$_9$ \cite{bhattacharyya2024imaging} and bilayer nickelate superconductors \cite{NV_LaNiO_2025NSR, NV_LaNiO_2025PRL} under high pressures. Despite these advances, the working pressure of NV centers is still not sufficient to study most hydride superconductors. For example, the superconducting transition temperature and required pressure for typical hydrides are:  H$_3$S (203 K @ 150 GPa) \cite{HS32015Nature},  CaH$_6$ (215 K @ 172 GPa) \cite{CaH2022PRL}, YH$_9$ (243 K @ 201 GPa) \cite{YH2021NC}. The record-high superconducting transition temperature of LaH$_{10}$ (250--260 K) is measured under pressures of 150--200 GPa \cite{LaH2019Nature, LaH2019PRL}. To investigate novel quantum states under ultrahigh pressures, it is essential to further increase the working pressure of diamond NV centers.

In this work, we show that the working pressure of NV centers can be extended to nearly 200 GPa, covering the pressures required  for most hydride superconductors. Using neon gas as the pressure transmitting medium, we experimentally demonstrate coherent quantum control and quantum sensing with NV centers in (111)-cut diamond anvil at pressures up to 191 GPa. Leveraging this capability, we conduct in-situ magnetic measurements of lanthanum hydride synthesized by laser heating lanthanum and ammonia borane (H$_3$NBH$_3$). Synchrotron X-ray diffraction (XRD) measurements indicate that the dominant phase in the sample is LaH$_{9.6}$ $Fm\overline{3}m$, and electrical resistance measurements yield a superconducting transition temperature of 216 K. By imaging the magnetic field distribution across the sample after zero-field cooling (ZFC) and under a uniform external magnetic field, clear signatures of Meissner repulsion are observed under 150 GPa. A comparison of ODMR spectra after field cooling (FC) and ZFC shows strong flux trapping in this sample. These results show that NV-based quantum sensing is a powerful tool for studying quantum materials under harsh conditions.
  %Notably, superconducting diamagnetism and localized magnetic field enhancement are observed simultaneously at some locations, indicating significant spatial inhomogeneity in the sample.

\subsection*{Quantum control at nearly 200 GPa}

\begin{figure}[t]
\centering
\includegraphics[width=1.0\textwidth]{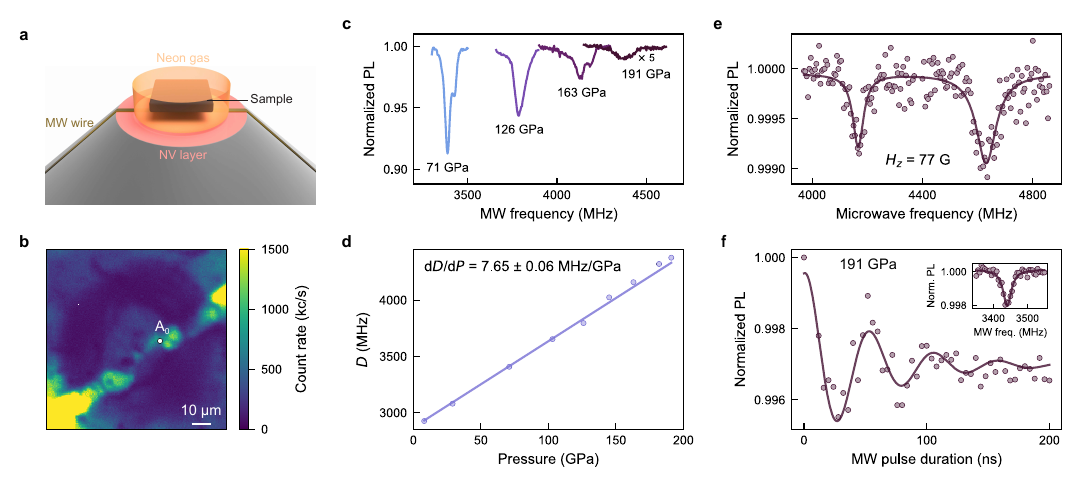}
\caption{\textbf{Quantum control of NV centers in diamond at nearly 200 GPa.}
    \textbf{a}, Schematic of the high-pressure chamber. A layer of NV centers is created on the diamond anvil by nitrogen ion implantation followed by high-temperature annealing. Optical excitation and fluorescence collection of the NV centers are performed through the transparent diamond window, while microwave pulses are applied via Pt wires placed on the culet. Neon gas is used as the pressure transmitting medium.
    %to maintain hydrostatic pressures. 
   % (111)-cut diamond is used so the pressure is applied along one of the NV axis.
    \textbf{b}, Confocal image of the diamond culet at $P$ = 145 GPa; the measured point is marked as A$_0$.
    \textbf{c}, Zero-field  optically detected magnetic resonance (ODMR) spectra at different pressures.
    \textbf{d}, Zero-field splitting, \textit{D}, as a function of pressure. The values of \textit{D} are extracted by fitting the zero-field ODMR spectra with single or double Lorentzian functions. Pressures are calibrated using the diamond Raman signal (for $P \ge $ 70 GPa) or the known \textit{D-P} relation (for $P<$ 70 GPa) \cite{NVHP2024NC}.
    \textbf{e}, ODMR spectrum at 191 GPa and $\textit{H}_z$ = 77 G.
    \textbf{f}, Rabi oscillations of NV centers at 191 GPa and $\textit{H}_z$ = 301 G. Inset: ODMR spectra under the same conditions. 
        }\label{Fig.1}
\end{figure}

Our primary goal is to extend the working pressure range of NV centers in diamond. To create and maintain high pressures up to 200 GPa, DACs with small culets (60 $\mu$m in diameter) are used. A layer of shallow NV centers is created on the (111)-cut diamond anvils through nitrogen ion implantation (20 keV and 2 $\times$ 10$^{14}$ cm$^{-2}$ dose) and followed by annealing under vacuum, as shown in Fig. \ref{Fig.1}a. Previous studies have shown that a uniform and hydrostatic pressure environment is crucial for maintaining the optical readout mechanism of NV spins \cite{NVHP2022CPL, NVHP2023PRB, NVHP2024NC, bhattacharyya2024imaging}. Therefore, we use neon-gas as the pressure transmitting medium (PTM) in the experiment, as it provides better hydrostatic pressure conditions than solid PTMs. The experiments are performed on a home-built ODMR setup; details of the experimental setup and methods are provided in Methods.

%Confocal fluorescence images and ODMR spectra are acquired at room temperature and under various pressures. The experimental pressures are calibrated using the Raman signal of diamond or the known $P-D$ relation of NV centers in (111)-cut diamond \cite{NVHP2024NC} (See Methods and Fig. S1 for details). 
Figure \ref{Fig.1}b shows a typical confocal image of the diamond culet after compressing the DAC1 sample to 145 GPa; the fluorescence from the NV centers, the platinum wires, and the edge of the diamond culet can be clearly distinguished. Figure \ref{Fig.1}c presents typical ODMR spectra of NV centers at different pressures. Although broadening and additional splitting appear at higher pressures, the ODMR contrast remains visible up to 191 GPa, setting a new record for the optical polarization and readout of NV spins under high pressures. By fitting the spectra with single or double Lorentzian functions, the zero-field splitting \textit{D} is extracted and plotted in Fig. \ref{Fig.1}d. This parameter increases linearly with pressure (coefficient = 7.65 $\pm$ 0.06 MHz/GPa), and can therefore serve as an independent criterion for pressure calibration. When an external magnetic field $\textit{H}_z$ is applied, the ODMR spectrum exhibits two distinct resonances (Fig. \ref{Fig.1}e), resulting from the Zeeman splitting between the $|m_s = \pm 1\rangle$ states of the NV centers. The external magnetic field can effectively suppress ODMR broadening caused by the pressure gradient. A sharp ODMR signal with a width of about 29 MHz is obtained under an external field of 301 G (inset of Fig. \ref{Fig.1}f). Under the same conditions, we measure the Rabi oscillations of NV centers, as shown in Fig. \ref{Fig.1}f, demonstrating that full quantum control is achievable at high pressures near 200 GPa.
%This observation demonstrates our ability to coherently manipulate ensembles of NV centers at this record-high pressure. In summary, these results serve as the basis for our later investigation of magnetism in LaH$_{x}$.

\subsection*{Synthesis and zero-resistance of LaH$_{9.6}$}

%Motivated by the improved full quantum control under ultrahigh pressure, the trapping of magnetic flux in lanthanum superhydride is investigated herein. 
We then proceed to synthesize and characterize lanthanum superhydride at high pressures.
Two lanthanum superhydride samples (DAC2 and DAC3) are synthesized by laser heating lanthanum and ammonia borane at approximately 150 GPa. The generation of H$_2$ and the decrease in pressure after laser heating indicate that chemical reactions occur in the sample chamber (Fig. S3). The H$_2$ also provides a suitable pressure environment for conducting ODMR measurements under high pressures, similar to the neon gas in the first experiment. 

\begin{figure}
\centering
\includegraphics[width=0.95\textwidth]{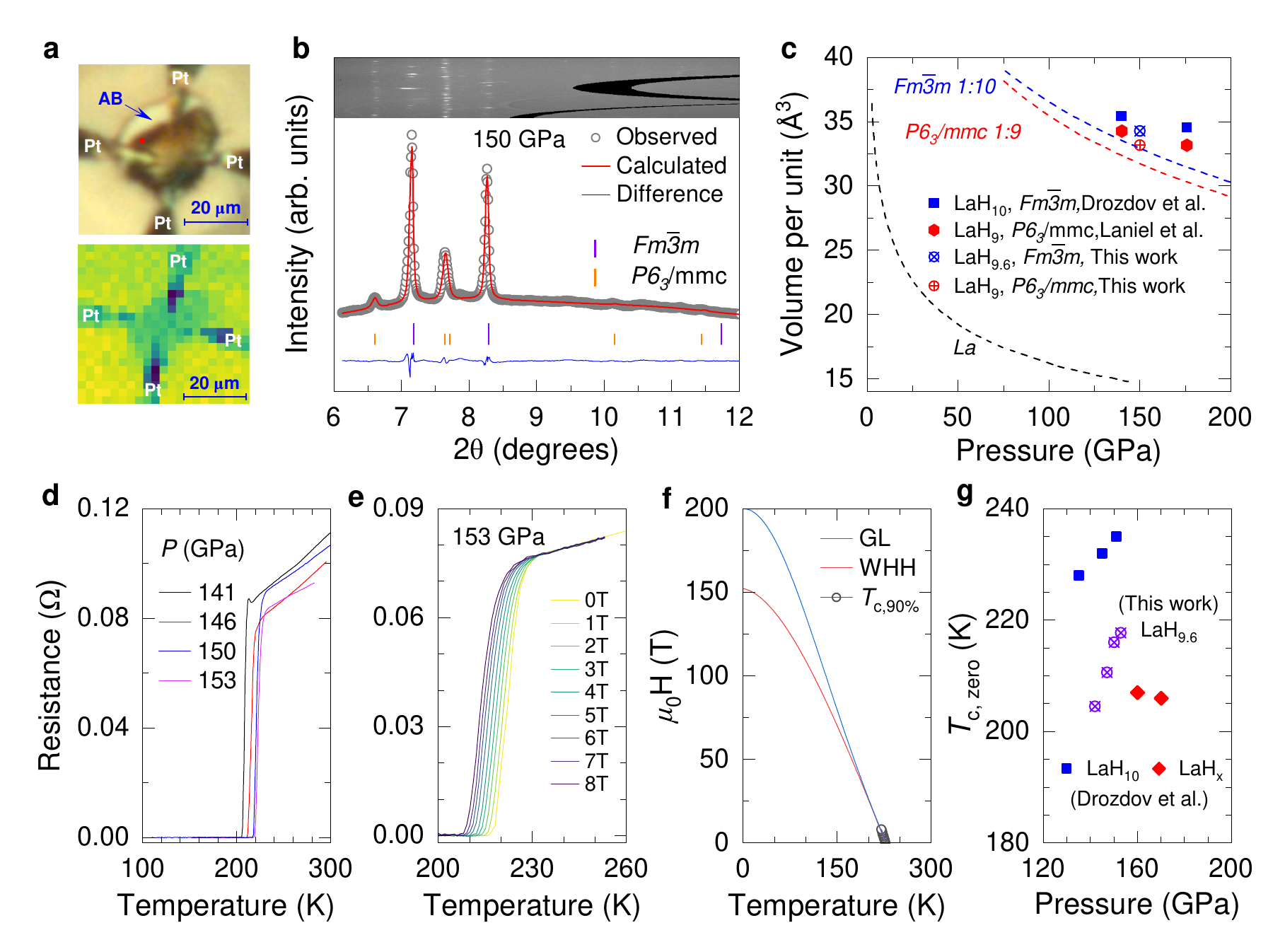}
\caption{\textbf{Synthesis and superconductivity of LaH$_{9.6}$.}
    \textbf{a}, Optical image of the sample chamber after laser heating (top) and the 2D synchrotron X-ray diffraction image of lanthanum superhydride at around 150 GPa (bottom). 
    \textbf{b}, XRD data collected at the red point in Fig. \ref{Fig.2}a along with the model fitting for the phases of $Fm\overline{3}m$-LaH$_{9.6}$ and $P6_3/mmc$-LaH$_{9}$. The experimental data points, calculated values, and Bragg peak positions are shown as small open circles, thin curves, and vertical sticks, respectively. The top panel shows the cake view of the raw X-ray diffraction patterns. 
    \textbf{c}, Pressure dependent unit cell volume per La atom for the different lanthanum superhydrides. The dashed blue, red and black lines indicate the $P-V$ relation of $Fm\overline{3}m$-LaH$_{10}$, $P6_3/mmc$-LaH$_{9}$ and $Fmmm$-La, respectively. The experimental results for the LaH$_{10}$, LaH$_{9}$ and LaH$_{9.6}$ are shown as blue and red dots.
    \textbf{d}, Superconducting transitions at different pressures.  
   \textbf{e}, Temperature dependence of the resistance at applied magnetic fields from 0 to 8 T at a pressure of around 153 GPa. 
   \textbf{f}, Upper critical field $H_\text{c2}$ as a function of temperature, fitted with the GL (blue) and WHH model (red).
   \textbf{g}, Pressure dependence of $T_\text{C,zero}$ for LaH$_{9.6}$ synthesized in this work compared with that of LaH$_{10}$ and LaH$_{x}$ reported in ref. \cite{LaH2019Nature}. 
              }\label{Fig.2}
\end{figure}

Synchrotron XRD measurements are conducted to determine the structure of the synthesized products in DAC2. As shown in Fig. \ref{Fig.2}(a--b), the sample can be indexed by the space groups of $Fm\overline{3}m$ and $P6_3/mmc$ depending on the structural features reported in similar systems in  literature \cite{LaH2019Nature,laniel2022high}. Based on the diffraction peak intensities, $Fm\overline{3}m$ is the primary phase present in the DAC2 sample (Fig. S4). The fitted lattice parameters and volumes at selected locations in DAC2 are summarized in Table S1. Although hydrogen occupancy cannot be determined experimentally due to its poor X-ray scattering ability, its concentration can be inferred by comparing the unit cell volume \cite{ji2019ultrahigh,chen2020superconductivity}. The calculated hydrogen contents are about 8.35-9.20 for the $P6_3/mmc$ phase and 9.23-9.75 for the synthesized $Fm\overline{3}m$ phases (Table S1). The refined volumes of $Fm\overline{3}m$ and $P6_3/mmc$ collected at the red point  (Fig. \ref{Fig.2}a) are shown in Fig. \ref{Fig.2}c, compared with the previously reported  $P-V$ results of LaH$_{10}$ and LaH$_{9}$\cite{LaH2019Nature,laniel2022high}. The calculated hydrogen concentration of $Fm\overline{3}m$ in this work shows a small deviation from the ideal value of 10, possibly due to the formation of hydrogen vacancies during the synthesis process.

Superconductivity in the synthesized lanthanum superhydride is confirmed by electrical transport measurements. The temperature-dependent resistance data at different pressures and under various external magnetic fields are shown in Fig. \ref{Fig.2}d and \ref{Fig.2}e, respectively. Sharp superconducting transitions and zero resistance are clearly observed at different pressures. These phenomena indicate the observed superconductivity arises from a single phase. In Fig. \ref{Fig.2}e, the resistance gradually shifts to lower temperatures, and the superconducting transition broadens under strong magnetic fields. The upper critical field at 0 K is extrapolated to be 156 and 200 T with the Ginzburg-Landau (GL) and the Werthamer-Helfand-Hohenberg (WHH) model fitting \cite{ginzburg2009theory, werthamer1966temperature}, respectively (Fig. \ref{Fig.2}f). Figure \ref{Fig.2}g summarizes the extracted $T_\text{c zero}$ of our sample, along with LaH$_{10}$ and LaH$_{x}$ reported by Drozdov et al. \cite{LaH2019Nature}. The large $T_\text{c}$ difference between our sample and LaH$_{10}$ indicates that the superconductivity is originates from a new phase. Considering the unstable superconductivity in $P6_3/mmc$-LaH$_{9}$ and the observed sharp superconducting transitions, we conclude that the superconducting phase in our sample is $Fm\overline{3}m$, LaH$_{9.6}$.
This phase is very stable and does not change during measurements lasting more than three months. We note that superconductivity in a similar phase (labeled as LaH$_{x}$) has been reported previously \cite{LaH2019Nature}.

\subsection*{Magnetic screening and flux trapping}

\begin{figure}
\centering
\includegraphics[width=0.9\textwidth]{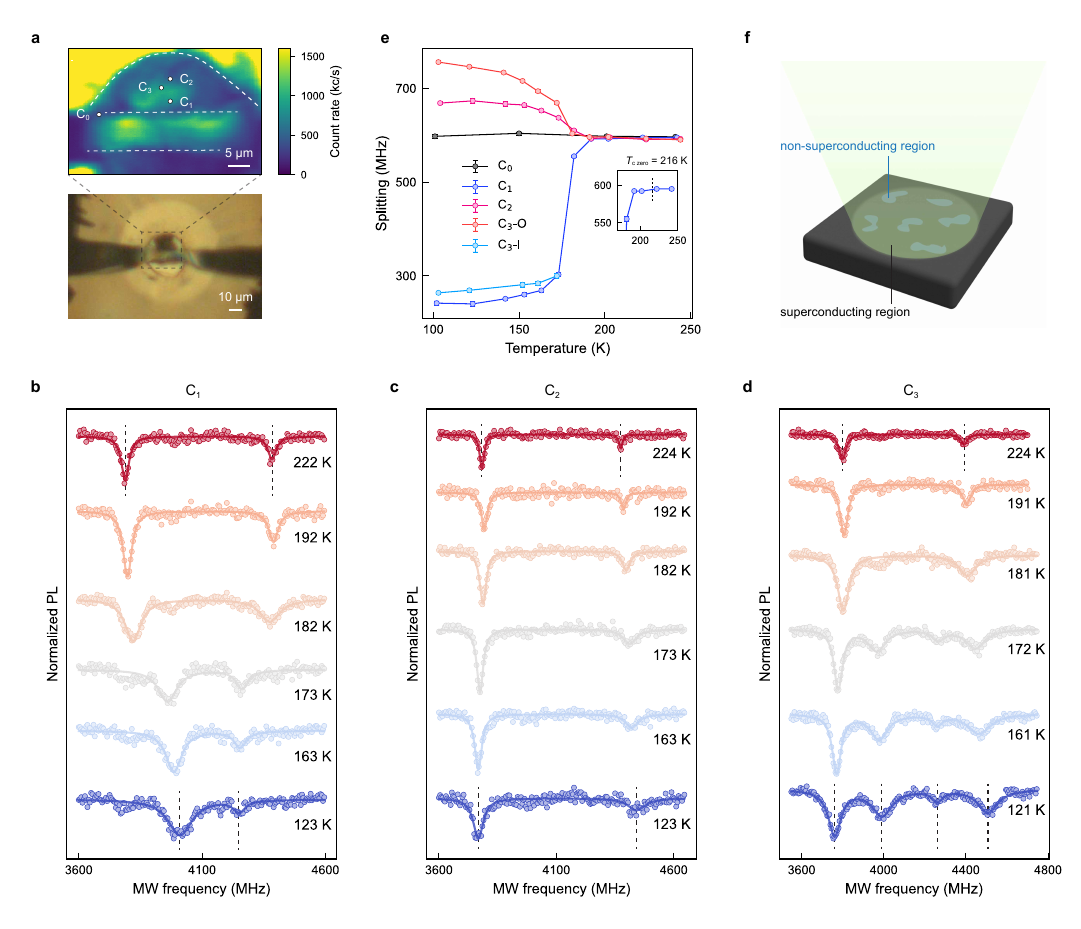}
\caption{\textbf{Magnetic screening of LaH$_{9.6}$ at 150 GPa.}
    \textbf{a}, Confocal fluorescence image (top) and bright-field image (bottom) of the LaH$_{9.6}$ sample.
    \textbf{b--d}, ODMR spectra of C$_1$--C$_3$ measured during field heating at $\textit{H}_z$ = 105 G after zero-field cooling. The solid lines are fits with multiple Lorentzian functions. At low temperature, ODMR spectra at C$_1$ (sample center) and C$_2$ (near the sample edge) reveal strong demagnetization and concentration of magnetic flux, respectively. Most  other test points (C$_3$ and more data in Fig. S6) show both effects simultaneously--four dips are observed at low temperature, and ODMR splitting extracted from the outer (inner) dips decreases (increases) with temperature. At high temperature, all ODMR spectra display a two-tip feature with identical splitting determined by the external magnetic field.  
    \textbf{e}, ODMR splittings of the examined points as a function of temperature. The outer and inner pairs of dips in the ODMR spectra of C$_3$ are labeled C$_3$-O  and C$_3$-I, respectively.
    \textbf{f}, Schematic illustrating the coexistence of superconducting and non-superconducting regions in the laser-focused area, which produces four-dip ODMR spectra when NV centers with a single orientation are used for measurement.
        }\label{Fig.3}
\end{figure}

We then characterize the Meissner effect of the lanthanum hydride sample using diamond quantum sensors. As shown in Fig. \ref{Fig.3}a, the sample region in DAC3 is identified by comparing the confocal image with the bright field image. In the zero-field cooling and field-warming (ZFC-FW) experiment, the sample is first cooled to 100 K at zero field. ODMR spectra are then acquired under a uniform external magnetic field. Typical ODMR spectra and their temperature dependence are shown in Fig. \ref{Fig.3}(b--d). For NV centers far from the sample, such as the C$_0$ position, a clear two-dip feature with barely temperature dependence is observed (Fig. S5). Therefore, ODMR spectra at this position serve as a reference to estimate the strength of the external magnetic field, which is fixed at 105 G. For NV centers directly below the sample, such as at the C$_1$ position, a much smaller splitting (242 MHz) is observed, indicating strong magnetic field screening from the sample. As temperature increases, a sharp transition of the ODMR splitting, which is linearly proportional to the strength of the local magnetic field, is observed. The transition starts at around 170 K, and completes at around 220 K (Fig. \ref{Fig.3}e inset). In contrast, for NV centers near the sample edge (C$_2$), ODMR splittings are larger at temperatures below $T_\text{c}$, and gradually decrease to the reference value (Fig. \ref{Fig.3}c,e). The initially larger splitting is attributed to the concentration of magnetic flux repelled from the superconducting regions. At temperatures above 220 K, ODMR splittings at different positions converge to nearly the same value.

\begin{figure}
\centering
\includegraphics[width=0.9\textwidth]{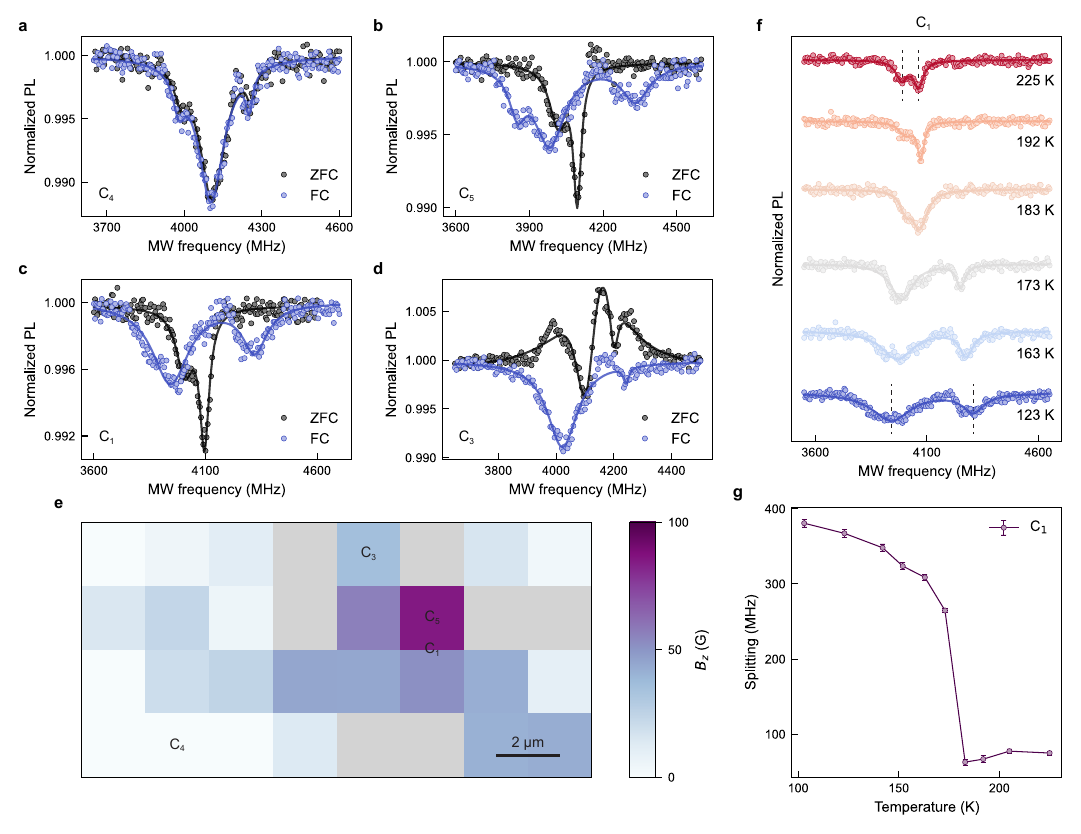}
\caption{\textbf{Imaging flux trapping using in-situ quantum sensors.}
    \textbf{a--d}, Typical zero-field ODMR spectra (of C$_4$, C$_5$, C$_1$ and C$_3$) measured after ZFD (black) and FC at $\textit{H}_z$ = 105 G (blue). The solid lines are fits with multiple Lorentzian functions. For NV centers at C$_4$, the two ODMR signals are nearly identical, indicating no flux trapping at this position. In contrast, most other measured points exhibit very different zero-field ODMR spectra after FC, indicating strong local flux trapping at these positions. The strength of the residual magnetic field is estimated from the ODMR splitting (see Methods).
    \textbf{e}, Spatial map of the residual magnetic field $B_z$. The center of the sample shows strong flux trapping. Positions with finite trapped field $B_z$, whose magnitude cannot be estimated, are marked in gray (Fig. S8).
    \textbf{f}, ODMR spectra of C$_1$ measured during zero-field heating after field cooling at $\textit{H}_z$ = 105 G.
    \textbf{g}, ODMR splittings of C$_1$ as a function of temperature.
        }\label{Fig.4}
\end{figure}

An interesting finding in this sample is that many of the examined points display a four-dip feature in their ODMR spectra, as shown in Fig. \ref{Fig.3}d and Fig. S6. Since the experiment is conducted at a high pressure of 150 GPa, only NV centers in the (111) orientation can survive \cite{bhattacharyya2024imaging}, and a specific magnetic field splits the resonance dip into two. The four-dip feature at low temperatures indicates that the NV centers in the detection volume experience two primary magnetic fields. Tracking the temperature dependence of these four dips reveals that the inner pair exhibits local diamagnetism, while the outer pair behaves similarly to C$_2$ (Fig. \ref{Fig.3}e). The four-dip ODMR spectra and their temperature dependence suggest that both superconducting and non-superconducting regions exist in the detection volume and contribute to the ODMR signals, as illustrated in Fig. \ref{Fig.3}f. The non-superconducting regions can be attributed to sample inhomogeneities introduced during the synthesis of lanthanum hydride by laser heating, consistent with the results of XRD measurements.

While clear local diamagnetism was observed, more compelling evidence of superconductivity is magnetic flux trapping. To probe the trapped flux, we applied a 105 G magnetic field, cooled the sample to 100 K, and then quenched the field to zero. A rectangular region over the sample is selected for ODMR measurements (Fig. S6). A second round of measurement is also performed after ZFC for comparison. Typical ODMR spectra are shown in Fig. \ref{Fig.4}a--d. At C$_4$, the ODMR spectra from FC and ZFC are nearly identical, indicating that no flux trapping occurred at this position (Fig. \ref{Fig.4}a). In contrast, the ODMR spectra at C$_5$ and C$_1$ (and many other positions) after FC show significantly larger splitting compared to their ZFC counterparts, as shown in Fig. \ref{Fig.4}b,c). The broadened resonant dips indicate strong magnetic field gradients at these positions. To further verify that the residual magnetic field results from flux trapping in the superconducting sample, we gradually warm the sample and measure ODMR spectra at C$_1$, as shown in Fig. \ref{Fig.4}f. The ODMR splitting decreases as temperature increases and eventually returns to the intrinsic value, indicating that the trapped flux is released (Fig. \ref{Fig.4}g). The abrupt change occurs at approximately 180 K, consistent with the transition window observed in the ZFC-FW measurement, as shwon in Fig. \ref{Fig.3}e.

The high spatial resolution of diamond quantum sensing enables us to image the flux trapping under megabar pressures. To quantify the trapped flux, all the measured ODMR spectra are fitted with multiple Lorentzian functions. The outermost pair of dips is used to estimate the strength of the residual magnetic field due to flux trapping. The axial trapped field $B_z$ is obtained by subtracting (in quadrature, see Methods) the ZFC splitting from the FC splitting. For certain points, such as C$_3$, the ODMR spectrum displays an additional positive envelope (Fig. \ref{Fig.4}d), possibly due to non-axial stress at the location. Nevertheless, their magnetic field dependence is similar to that of the normal ODMR spectra (competition between the Zeeman effect and transverse stress; Fig. S7). The results from all measured points provide a spatial map of the axial trapped field $B_z$ (Fig. \ref{Fig.4}e). These results further confirm that the magnetic signatures we observed are indeed due to superconductivity.

\subsection*{Discussion and Conclusions} 
In summary, our results demonstrate that NV-based quantum sensing provides indispensable experimental evidence of magnetic field screening and flux trapping in lanthanum hydride superconductors under megabar pressures.  Compared to conventional probes \cite{minkov2023magnetic}, NV centers offer spatially resolved demagnetization signals of the lanthanum
hydride sample, enabling the investigation of sample   inhomogeneities at the microscale, which is crucial for  optimizing the synthesis of hydride superconductors under high pressures. Using H$_2$ as the hydrogen source \cite{LaH2019Nature}, improved hydrostatic pressure conditions and the Meissner effect at higher temperatures can be expected in the near future .
%In particular, even for superconducting samples with good zero resistance, spatial resolved characterization is necessary in revealing its inhomogeneities at microscle. 

Our work demonstrates the functionality of NV quantum sensors at pressures up to 200 GPa, with significant potential to further increase the working pressure of NV centers in diamond. Pressures higher than 300 GPa have been achieved in experiments \cite{HP_2018RMP, HP_500GPa}. To enable NV centers to operate under such extreme conditions, substantial effort is required to optimize optical excitation and collection, particularly to address the weaker fluorescence signal of NV centers and the stronger background from diamond itself under ultrahigh pressures. Looking forward, the combination of high pressure with other conditions, such as strong magnetic fields (10 T), ultra-low temperatures (100 mK), or high temperatures (2000 K), offers new opportunities to discover and understand novel quantum states under synergetic extreme conditions. By leveraging advanced quantum control and quantum properties of NV centers \cite{NV_Squeezing_2025, NV_Nature2025, Correlation2019PRL, NV_Review_2024NRP}, practical quantum advantage can be achieved under extreme conditions.

\subsection*{Methods}

\textbf{NV fabrication and sample loading.}
Three nonmagnetic symmetric DACs (DAC1--3) were prepared to generate high pressures. The pressure was determined from the first-order Raman spectra of the diamond anvils \cite{akahama2005raman, akahama2010pressure}. The diamonds used in the DACs had 50, 60 or 80 $\mu$m culets and were beveled at 8--8.5$^\circ$ to a diameter of about 300 $\mu$m. The top anvils of both DAC1 and DAC3 were (111)-cut diamonds (type \Romannum{1}b for DAC1 and type \Romannum{2}a for DAC3), which were implanted with a layer of NV centers. To create the NV layer, the diamonds were implanted with N$^+$ ions at 20 keV and a dose of 2 $\times$ 10$^{14}$ cm$^{-2}$ and subsequently annealed at \SI{800}{\celsius} for 2 h. All the other anvils were (100)-cut diamonds. In DAC1, a 40 $\mu$m hole was laser drilled to serve as the sample chamber and then prefilled with a mixture of cubic boron nitride (cBN) and epoxy to electrically isolate the rhenium gasket. A thin platinum foil (1 $\mu$m thick $\times$ 5 $\mu$m wide) was positioned on the diamond culet for microwave transmission. Finally, neon gas was loaded into the sample chamber as the pressure transmitting medium to provide better hydrostatic conditions. \\

\noindent \textbf{LaH$_{9.6}$ synthesis.} For DAC2 and DAC3, samples were loaded in an argon-filled glovebox (O$_2$ and H$_2$O $<$ 0.1 ppm) to avoid air exposure. A 1--2 $\mu$m thick La piece was placed in direct contact with ammonia borane (H$_3$NBH$_3$), which served as both the hydrogen source and the pressure transmitting medium \cite{CeYH9_2024NC}. A similar platinum foil was positioned as described above in DAC1. After sealing, the DAC was compressed in multiple steps to the target pressure required to synthesize lanthanum hydrides. The sample was then laser-heated from one-side to 1500--2000 K for 1--2 minutes using a yttrium-aluminum-garnet (YAG) laser. A laser power of 15 W was used and focused onto a spot of approximately 5 $\mu$m. \\

\noindent \textbf{ODMR setup.} 
ODMR measurements were performed on a home-built confocal microscope system equipped with a commercial cryostat (s100, Montana Instruments). A 532-nm laser (CNI Laser) was focused by a long working distance objective (LT-APO/ULWD/VISIR/0.35, attocube) to initialize the NV centers. The emitted fluorescence was filtered by a 650-nm longpass filter (Thorlabs) and detected using a single-photon counting module (SPCM-780-10-FC, Excelitas). Laser and microwave pulses were generated with an acousto-optic modulator (Gooch \& Housego) and an RF switch (Mini-circuits), respectively. Pulse synchronization and counter timing were controlled by a programmable multi-channel pulse generator (PBESR-PRO 500, SpinCore). The ODMR spectra were recorded in a continuous-wave mode, with laser excitation, microwave driving, and photon counting performed  simultaneously while sweeping the microwave frequency.
\\

\noindent \textbf{Electrical transport measurements.} 
A typical symmetric DAC (DAC2) made from Cu-Be alloy was used for electrical transport measurements at external magnetic fields ranging from 0 to 8 T. For the resistance measurements, the rhenium gasket was insulated with a mixture of cubic boron nitride (cBN) and epoxy. Four platinum electrodes were attached in the chamber to connect the sample to external copper wires using the Van Der Pauw method for resistance measurements at high pressures \cite{philips1958method}. The measurements were conducted in a commercial cryostat equipped with a Keithley 6221 current source and a 2182A nanovoltmeter.\\

\noindent \textbf{Synchrotron X-ray diffraction  measurements.} At high pressures, X-ray diffraction patterns were collected at the ID27 synchrotron beamline at the European Synchrotron Radiation Facility (Grenoble, France) using the EIGER2 X CdTe 9M detector. The wavelength was set to 0.3738 \AA{} and the spot size was 700 nm. X-ray diffraction patterns were collected with DAC2. The obtained two-dimensional XRD patterns were integrated into one-dimensional profiles using the DIOPTAS software \cite{prescher2015dioptas}. The data were then analyzed with the JANA software based on the LeBail method \cite{le2005whole}.\\

\noindent\textbf{Calculation of the axial trapped field $\textit{B}_z$.} 
In the absence of external disturbances, the ground-state Hamiltonian of an NV center is given by $H_0 = DS_z^2$. Here, $D \approx (2\pi)$ $\times$ 2.87 GHz is the zero-field splitting parameter at ambient conditions, and $\vec{S} = (S_x, S_y, S_z)$ is the spin-1 operator expressed in the local frame of the NV center ($z$ is defined along the NV axis). A magnetic field $\vec{B}$ couples to the NV center through the Hamiltonian $H_B = \gamma_B \vec{B} \cdot \vec{S}$, where $\gamma_B = (2\pi)$ $\times$ 2.802 MHz is the gyromagnetic ratio of the NV center \cite{barry2020sensitivity}. Stress couples via the Hamiltonian $H_s = \Pi_zS_z^2 + \Pi_x(S_y^2 - S_x^2) + \Pi_y(S_xS_y + S_yS_x)$, where $\Pi_x$, $\Pi_y$ and $\Pi_z$ are functions of the stress tensor $\boldsymbol{\sigma}$ \cite{bhattacharyya2024imaging, hsieh2019imaging} Stress components that break the $C_{3v}$ symmetry of the NV center induce a splitting $2\Pi_\perp = 2\sqrt{\Pi_x^2 + \Pi_y^2}$. The Zeemann splitting $2\gamma_BB_z$ from an axial magnetic field $B_z$ adds in quadrature with the stress-induced splitting $2\Pi_\perp$. This results in a total ODMR splitting of $\Delta = \sqrt{(2\gamma_BB_z)^2 + (2\Pi_\perp)^2}$, from which $B_z$ can be extracted. It can be shown that, even in the presence of $B_x$ or $B_y$ of similar magnitude, the field-induced splitting is still dominated by $B_z$. Therefore, we continue to use the above equation to approximate $B_z$ for simplicity. Specifically, the ODMR splittings were extracted by fitting the spectra with multiple Lorentzian functions. For spectra with more than two dips, the outermost pair of dips was used, as it corresponds to the highest typical $B_z$ value.

\subsection*{Data availability}
The data that support the findings of this study are available from the corresponding author upon reasonable request.
%The published data of this study are available on the Zenodo public database.

\newpage
\bibliography{references}

\subsection*{Acknowledgements}
This work is supported by the National Key Research and Development Program of China (Grant Nos. 2023YFA1608900, 2021YFA1400300, 2024YFA1611300), the Innovation Program for Quantum Science and Technology (Grant No. 2023ZD0300600), the National Natural Science Foundation of China (Grant Nos. 12022509, T2121001, 12074422, 12374468, 12375304, 12404163), and the Chinese Academy of Sciences (Grant No. YSBR-100). We thank the high-pressure synergetic measurement station (A9) of the Synergtic Extreme Condition User Facility (SECUF) for assistance with high-pressure resistivity measurements and laser heating. We thank Yue Xu and Rui-Zhi Zhang for fruitful discussions.

\subsection*{Author contributions}
Y.C., and Z.-X.H. performed the ODMR measurements and analyzed the data. J.W. and L.C. synthesized the LaH$_{9.6}$ sample, and prepared the electronic and Raman measurements. J.-W. F fabricated shallow NV centers on the DACs. J.C. and H.G. performed the XRD measurements. G.-Q.L., X.Y., and L.C. supervised the project. Y.C., L.C., and G.-Q.L. wrote the paper with input from all authors. All authors commented on the manuscript.

\subsection*{Competing interests}
The authors declare no competing interests.

\end{document}